\documentclass[%
 reprint,
superscriptaddress,
groupedaddress,
 amsmath,amssymb,
 aps,
]{revtex4-2}

\usepackage{graphicx}
\usepackage{dcolumn}
\usepackage{bm}

\newcommand{\etal}{\textit{et al.}~}
\newcommand{\reffig}[1]{Fig.~\ref{#1}}
\newcommand{\eq}[1]{Eq.~\eqref{#1}}

\begin{document}

\preprint{APS/123-QED}

\title{Skipping the boundary layer: high-speed droplet-based immunoassay using Rayleigh acoustic streaming}

\author{Qi Wang}
\affiliation{%
	ASIC and System State Key Laboratory, School of Microelectronics, Fudan University, Shanghai 200433, China
}%

\author{Zhe Ding}%

\affiliation{Viral Hemorrhagic Fevers Research Unit, CAS Key Laboratory of Molecular Virology and Immunology, Institut Pasteur of Shanghai, Chinese Academy of Sciences, Shanghai 200031, China}
\affiliation{University of Chinese Academy of Sciences, Beijing 100049, China}

\author{Gary Wong}%

\affiliation{Viral Hemorrhagic Fevers Research Unit, CAS Key Laboratory of Molecular Virology and Immunology, Institut Pasteur of Shanghai, Chinese Academy of Sciences, Shanghai 200031, China}

\author{Jia Zhou}
\affiliation{%
	ASIC and System State Key Laboratory, School of Microelectronics, Fudan University, Shanghai 200433, China
}%
\altaffiliation{corresponding authors}
\author{Antoine Riaud}
\affiliation{%
	ASIC and System State Key Laboratory, School of Microelectronics, Fudan University, Shanghai 200433, China
}%
\altaffiliation{corresponding authors}

\email{antoine\_riaud@fudan.edu.cn,jia.zhou@fudan.edu.cn}

\date{\today}

\begin{abstract}
 Acoustic mixing of droplets is a promising way to implement biosensors that combine high speed and minimal reagent consumption. To date, this type of droplet mixing is driven by a volume force resulting from the absorption of high-frequency acoustic waves in the bulk of the fluid. Here, we show that the speed of these sensors is limited by the slow advection of analyte to the sensor surface due to the formation of a hydrodynamic boundary layer. We eliminate this hydrodynamic boundary layer by using much lower ultrasonic frequencies to excite the droplet, which drives a Rayleigh streaming that behaves essentially like a slip velocity. Three-dimensional simulations show that this provides a threefold speedup compared to Eckart streaming. Experimentally, we shorten a SARS-CoV-2 antibody immunoassay from 20 min to 40 s.   
 
\begin{description}
\item[Keywords] Rayleigh acoustic streaming,  efficient mixing, immunoassay
\end{description}
\end{abstract}

\maketitle


\section{Introduction}

Surface biosensors integrated in microfluidic devices enable the fast detection of molecules in small liquid samples. However, they face a trade-off between detection speed and sample waste. Indeed, the detection of dilute molecules by surface biosensors proceeds in three steps: the mass transfer of analyte from the bulk of the sample to the sensor surface, the adsorption of the analyte on the sensor surface (which can be made specific using antibody, DNA hybridization, molecular imprints) and the detection of the adsorbed analyte (by fluorescence, surface plasmon resonance (SPR), field-effect transistor...). Most often, mass transfer is the slowest step of all, and therefore determines the detection speed of the sensor \cite{gervais2006mass,hansen2012transient,pereiro2020shake}. This slow mass-transfer is mainly due to the formation of a concentration boundary layer that prevents the analytes from the bulk to reach the sensor. The thickness of this boundary layer can be decreased by introducing convection (mixing). However, for the simplest devices such as a biosensor embedded in a microchannel, increasing the flow rate also increases the waste of analyte \cite{gervais2006mass} and the resulting shear may also damage the sensor \cite{pereiro2020shake}.

The mixing of sessile droplets is a promising strategy to address this trade-off between detection speed and waste. On the one hand, convection ensures a negligible boundary layer thickness (fast mass transfer) and on the other hand, the batch-mode mixing ensures recycling of the outgoing fluid. Such droplets can be mixed using electrowetting actuation, electro-osmosis, Marangoni effect or acoustic streaming (AS), with the latter featuring high flow speed, insensitivity to fluid composition and viscosity, excellent biocompatibility and the ability to mix pinned droplets (which is not possible for electrowetting). 

Acoustic streaming results from the transfer of pseudo-momentum from an acoustic wave to the fluid, which can happen through two mechanisms \cite{vanneste2011streaming,sadhal2012acoustofluidics}: due to the viscous attenuation of the wave in the bulk (Eckart streaming \cite{eckart1948vortices}) and due to the ultrasonic shear near the droplet boundaries (Rayleigh streaming) \cite{rayleigh1884circulation,nyborg1958acoustic,bach2018theory}. Eckart streaming is easily generated using high-frequency surface acoustic waves (SAW) ultrasonic transducers (interdigitated transducers) and behaves like a volume force that can mix droplets \cite{shilton2008particle} and, therefore, accelerate surface bioassays such as DNA hybridization assays \cite{wixforth2003acoustically}, immunoassays \cite{bourquin2011integrated}, and SPR detection \cite{renaudin2010integrated}. However, while Eckart streaming can reduce the thickness of the concentration boundary layer by stirring the droplet, the no-slip boundary condition at the solid-liquid interface creates a slow-moving hydrodynamic boundary layer near the sensor surface. As a result, the mixing improvement is limited by the slow convection near the solid wall \cite{salman2007numerical,lebedev2004passive}. 

Unlike Eckart streaming, Rayleigh streaming behaves like a slip velocity that reaches an asymptotic value within a few micrometers from the surface \cite{rayleigh1884circulation,nyborg1958acoustic,bach2018theory}. In this regard, it obliterates the hydrodynamic boundary layer near the sensor, and therefore ensures a faster detection. In most studies, Rayleigh streaming is created by oscillating sharp edges \cite{li2019acoustofluidic,zhang2021mixing,rasouli2019ultra} or microbubbles \cite{chen2018multiplexed,meng2019sonoporation,marin2015three}, and therefore the slip boundary condition is limited to the surface of the edges and bubbles. Instead, when considering surface biosensing applications, the slip velocity should occur on the sensor to maximize the detection speed. Although this type of full-device Rayleigh streaming has been achieved in microfluidic channels with hard-boundaries \cite{bengtsson2004ultrasonic}, we are not aware of studies where it would have been applied to droplets in order to accelerate surface biosensing.


In this paper, we develop a microfluidic chip to expedite surface bioassays by mixing droplets without contact. Unlike previous studies based on Eckart streaming, using Rayleigh streaming eliminates the hydrodynamic boundary layer at the bottom of the droplet and accelerates analyte mass transfer to the sensor, which shortens the detection time. We first use finite element simulation to show that Rayleigh streaming is faster than Eckart streaming at equal flow velocity in the droplet. We then use this Rayleigh streaming to shorten immunoassays from 20 min to just 40 s. 

\begin{figure*}
	\includegraphics{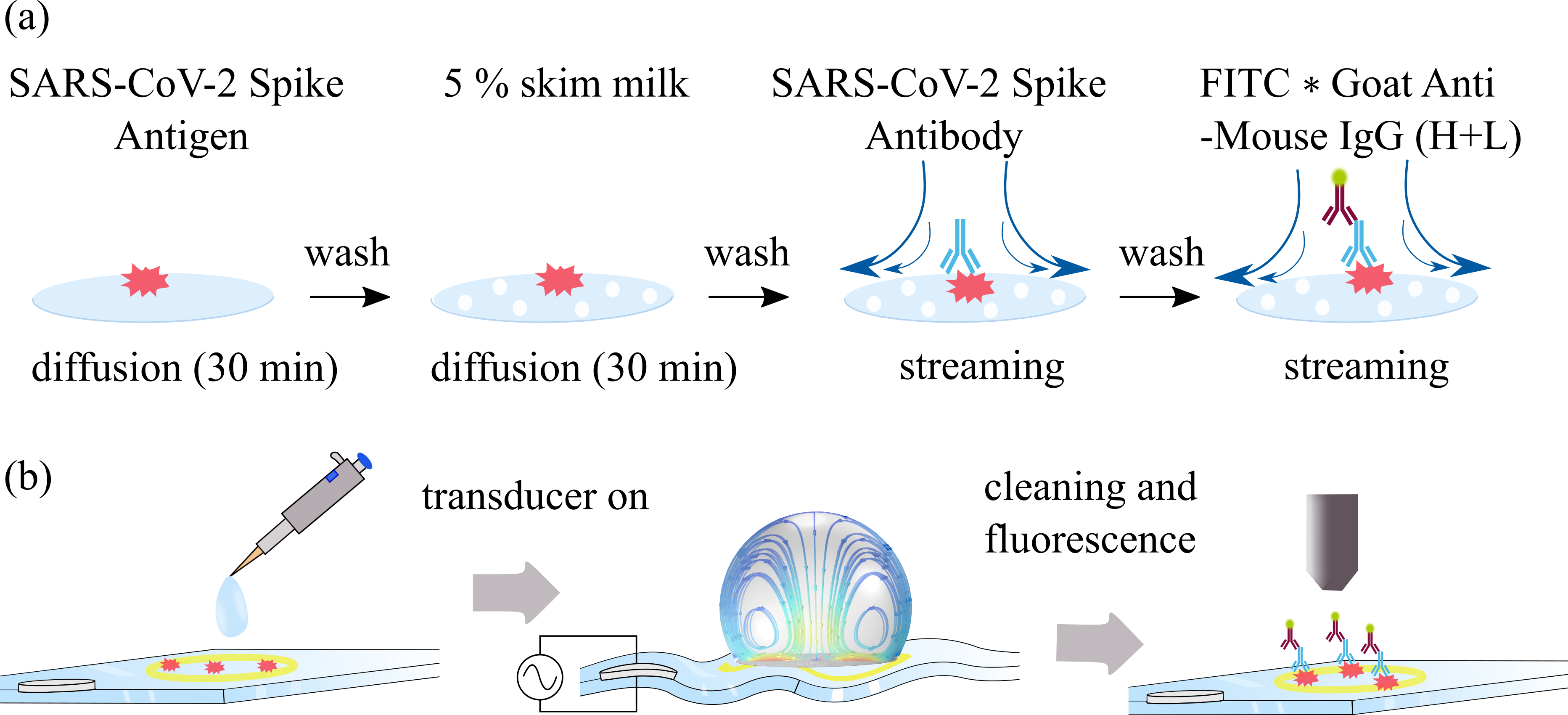}
	\caption{\label{fig:schematic diagram} Accelerated SARS-CoV-2 antibody detection immunoassay. (a) Main steps of the immunoassay: spike proteins of the virus are deposited on a treated glass substrate, then skim milk is added to prevent non-specfic adsorption. The analyte (primary antibody) can bind to the antigen is remains on the surface despite washing. Finally, the analyte is detected by a secondary fluorescent antibody. During the latest two stages, the mass transfer of antibodies to the sensor surface can be accelerated by mixing the liquid. (b) Method for the ultrasonic mixing: a droplet is deposited on the biosensor. When the transducer is turned on, the ultrasonic agitation of the glass is transmitted to the liquid, which triggers a steady acoustic streaming. After experiment, the glass slide is washed and the substrate fluorescence is measured by microscopy.}
\end{figure*}

\begin{figure*}
	\includegraphics{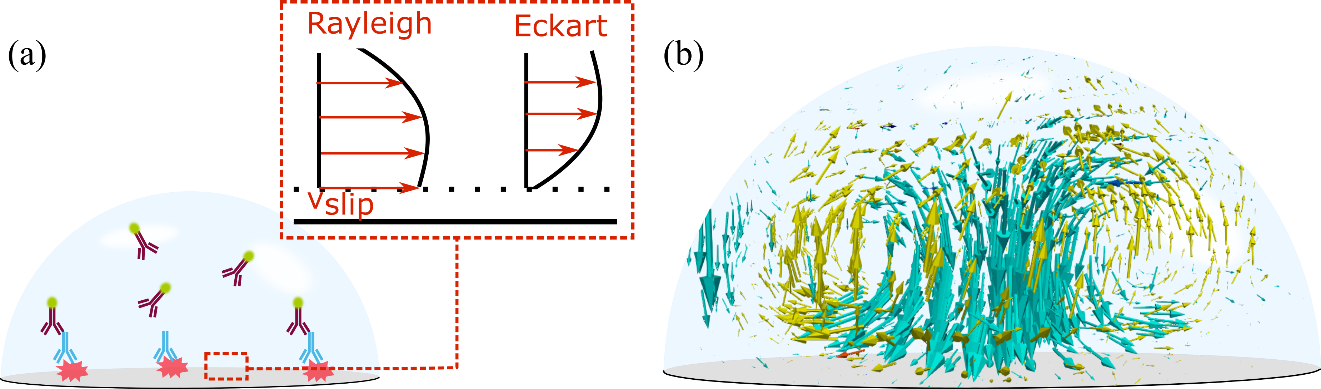}
	\caption{\label{fig:streaming GDPT} Acoustic streaming velocity fields. (a) Main difference between Eckart and Rayleigh streaming. Rayleigh streaming velocity is nonzero even within a few micrometers from the biosensor surface, whereas Eckart streaming features a hydrodynamic boundary layer due to the no-slip boundary condition on the sensor surface. (b) Experimental acoustic streaming field. To facilitate visualization, yellow and green arrows represent upward and downward velocities, respectively.  }
\end{figure*}

\section{Methods}
Our experiment (\reffig{fig:schematic diagram}(a)) aims to detect the presence of SARS-CoV-2 antibodies (thereafter referred as the analyte). The biosensor is first coated with SARS-CoV-2 spike proteins (the antigen). The analyte then binds with the recombinant spike protein from SARS-CoV-2, and is subsequently detected with a secondary antibody (FITC $\ast$ Goat anti-mouse IgG). The mass transfer of species to the surface can be accelerated by mixing the droplet with acoustic streaming (\reffig{fig:schematic diagram}(b)). A piezoelectric disc excited by an alternating voltage transmits its vibration to the droplet by the mean of a glass slide. Nonlinear interactions between the fluid and the solid then result in a steady flow in the droplet: the Rayleigh acoustic streaming.  Eventually, the droplet is rinsed and the adsorbed species are observed by fluorescence microscopy.

\subsection{Materials and Chemicals}
Fluorescence microparticles were purchased from Huge Biotechnology (China),
2019-nCoV (SARS-CoV-2) Spike (S) Protein (His-Avi) was purchased from Yazyme Biomedical Technology Co. (CG202-01; China), SARS-CoV-2 (2019-nCoV) Spike Neutralizing Antibody was purchased from Sino Biological (40591-MM43; China), FITC * Goat Anti-Mouse IgG(H+L) was provided by Beyotime Biotechnology (A0568; China), 3-APTES toluene solution was provided by Aladdin Bio-Chem Technology (A107147; China), Toluene and Methanol were provided by Sinopharm (10022818 and 10014108, respectively; China). All chemicals and reagents were used as received without further purification.

\subsection{Surface chemistry}
The glass slides are first functionalized with 3-aminopropyl triethoxysilane (3-APTES) to enable the binding of proteins. The functionalized substrate is then used either to measure the mass transfer of fluorescent antibodies (FITC $^{*}$ Goat Anti-Mouse lgG (H+L)) or to perform an immunoassay. 

\subsubsection{Preparation of APTES-functionalized glass slides}\label{sec: APTES}
A clean glass slide is first activated (hydroxilated) by treating it for 3 min with an air plasma. The glass slide is then silanized by immersion in a 20 mM 3-APTES toluene solution overnight. Finally, the substrate is rinsed with toluene and methanol to remove the unbounded 3-APTES. 

\subsubsection{Mass transfer assay}
A 0.2 mg/mL FITC $^{*}$ Goat Anti-Mouse lgG (H+L)) solution is directly incubated on the APTES-functionalized substrate (Section \ref{sec: APTES}). After a predetermined duration, the slide is rinsed thoroughly with DI water and observed by fluorescence microscopy. The exposure time is 307 ms.

\subsubsection{Fluorescent immunoassay}
The experiment can be divided in three steps: the substrate functionalization with APTES, priming with 2019-nCov antigen and the analyte detection itself. When acoustic streaming is used, only the analyte detection is accelerated.

\paragraph{Substrate priming}
The assay begins with a priming step where 0.1 mg/mL 2019-nCoV (SARS-CoV-2) Spike (S) Protein (His-Avi) are incubated for 30 minutes at 25 $^{\circ}$C on the APTES-functionalized glass substrate (Section \ref{sec: APTES}). Accordingly, the amount of spike protein used for each 2 $\mu$L droplet is 0.2 $\mu$g. In order to prevent non-specific adsorption of other proteins on remaining APTES sites, the substrate is then immersed in a solution of 5 \% skim milk diluted with 0.5 \% Tween Phosphate Buffer Saline (PBST) for 30 minutes at 25 $^{\circ}$C. 

\paragraph{Analyte detection}
Before the assay, the optimal antibody concentration for comparison between diffusion and acoustic mixing is determined by a pure diffusion titer assay described in SI. We find that the fluorescent immunoassay can detect antibody concentrations ranging from 2 nM to 60 nM of SARS-CoV-2 (2019-nCoV) Spike Neutralizing Antibody. A concentration of 28.5 nM was selected for our experiments. Primary antibody specificity, secondary antibody specificity and blank comparison results are shown in SI. 

In our experiments, the target analyte is suspended in a 2 $\mu$L droplet. Acoustic streaming mixing is active during the whole immunoassay. A solution of SARS-CoV-2 (2019-nCoV) Spike Neutralizing Antibody is first incubated on the substrate, then thoroughly rinsed with deionized (DI) water. The secondary antibody (0.2 mg/mL FITC $^{*}$ Goat Anti-Mouse lgG (H+L)) is then incubated on the substrate and rinsed thoroughly with DI water. Overall, 0.4 $\mu$g of secondary antibody is used for each droplet. To evaluate the amount of adsorbed antibody, fluorescence microscopy images of the sensor are recorded, and the average fluorescence intensity is computed with ImageJ. 


\subsection{Generation of acoustic streaming}
Rayleigh acoustic streaming is powered by the shearing motion of acoustic waves at the interface between a solid and a fluid. In our experiments, the acoustic power is provided by a piezoelectric transducer glued to the glass slide, and located several centimeters away from the droplet. The acoustic vibrations are then transmitted through the solid to the liquid, where they accumulate due to acoustic resonance.

\paragraph{Electroacoustic setup}
As shown in \reffig{fig:schematic diagram}(b), a lead zirconate titanate (PZT) piezoelectric disc is glued to the edge of a $50\times50\times1.8$ mm glass slide using a cyanoacrylates glue. To enhance repeatability, the disc was aligned with the glass slide using a 3D-printed template and pressed using a binder clip. The disc is powered by an AG1021 (T\&C) radiofrequency power amplifier connected to a signal generator (2207B, Picoscope). The excitation frequency was set to 800 kHz and the power output from the amplifier regulated to 2 W. During vibrometry experiments, the AG1021 was replaced with a LZY-22+ amplifier (Minicircuits) because the AG1021 could not be transported to the vibrometer. The signal generator voltage amplitude was set to 2 V, and the transmitted power was measured to be 4.63 W, 4.80 and 4.95 W at 800, 830 and 860 kHz, respectively.

\paragraph{Laser Doppler vibrometry}
The vibration amplitude of the droplet is measured by a Laser Doppler Vibrometer  (LDV) \cite{royer1986optical}. The laser hits the apex of the droplet and bounces back through a 40$\times$ microscope lens. Thanks to its large numerical aperture, such high-magnification lens can collect light from many directions which makes it relatively insensitive to small misalignment errors between the droplet vibrating surface and the vibrometer. The droplet vibration amplitude (120 nm) being comparable to the laser wavelength (633 nm), the perturbation approach described in Royer \etal \cite{royer1986optical} was not suitable. The non-perturbative method is described in SI. The PZT was controlled by amplifier Minicircuits LZY-22+ and Picoscope 2207B (2 V). During vibrometry characterization, the droplet profile was photographed by an auxiliary side-looking microscope.

\paragraph{Acoustic mixing assay}
During acoustic mixing assays, a 2 $\mu$L droplet is placed on the glass substrate (\reffig{fig:schematic diagram}). The droplet composition varies depending on the assay and is given in the corresponding method section (immunoassay/laser vibrometry/particle tracking). To enhance repeatability, the droplet contact line is immobilized using a gold ring patterned by photolithography and made hydrophobic using 1-dodecanethiol. In immunoassays, the gold ring was replaced with a silicone ring made of polydimethysiloxane. Once the droplet is in position, the transducer is activated and the droplet content is mixed using acoustic waves for the entire duration of the incubation step. Once the assay is completed, the surface is rinsed with DI water and observed using fluorescence microscopy.

\paragraph{General Defocusing Particle Tracking (GDPT)}
In order to visualize the acoustic streaming, the droplet is is seeded with 2 $\mu$m fluorescent particles (name?,brand?). An f=200.0 mm cylindrical lens (LJ1653RM-A, Thorlabs) is placed above the microscope objective to introduce a strong astigmatism that allows to calculate the out-of-plane location ($z$) of the particles based on their shape of their deformed image \cite{muller2013ultrasound,barnkob2015general,barnkob2021defocustracker,rossi2019interfacial} (see supplementary information for a detailed example). These calculations of the particle position and subsequent reconstruction of the particle trajectories across multiple video frames are done using the  Matlab implementation of the general defocus particle tracking (GDPT) method \cite{barnkob2015general}. 

\subsection{Simulation of acoustic streaming and mass transfer}
Compared to surface acoustic wave (SAW) driven Eckart streaming \cite{shiokawa1990study,raghavan2010particle,riaud2017influence}, Rayleigh streaming in sessile droplets has received a limited attention. Peng \etal \cite{peng2022concentration} have used a perturbation method to compute the acoustic streaming in two dimensions. In their paper, they resolve the viscoacoustic boundary layer to calculate the acoustic forcing. This requires a very fine mesh in order to capture all the details of this boundary layer \cite{muller2012numerical} and therefore this method is limited to two-dimensional models due to computer memory limitations. Here, we use the equivalent slip velocity as given by \eq{eq: Nyborg} to simulate acoustic streaming in three dimensions. For a given radial standing acoustic velocity field $\tilde{v}_r (t,r) = \tilde{V}_r(r)\sin(\omega t)$, Nyborg provides an expression of the slip velocity $\bar{v}_{slip}$ in the axisymmetric incompressible case \cite{nyborg1958acoustic}. In the appendix, we show that this expression is also valid in the compressible case:

\begin{equation}
	\bar{v}_{slip} = -\frac{3}{8\omega} \left. \left( \partial_r  {\tilde{V}_r}^2\right) \right|_\text{wall}  -\frac{1}{2 r \omega} \left. \left( {\tilde{V}_r}^2\right) \right|_\text{wall}.
	\label{eq: Nyborg}
\end{equation}
The simulations are implemented with Comsol multiphysics using two-dimensional axisymmetric finite element models, using laminar flow (creeping flow) and mass transfer of dilute species modules. First, the acoustic field in the droplet is estimated to the LDV calibration (see SI), and the associated velocity field $\tilde{V}_r(r)$ is input in the model. The acoustic streaming is then computed from \eq{eq: Nyborg} using a stationary solver for the laminar flow equations. Next, this streaming field is used as input for a time-dependent convective mass-transfer and reaction problem. In the latter, antibody adsorption is simulated as a fast irreversible reaction. In first approximation, a typical antibody diffusivity is taken as $4\times10^{-11}$ m$^2$/s \cite{pokric1979two}.

\section{Results and discussion}

\begin{figure*}
	\includegraphics{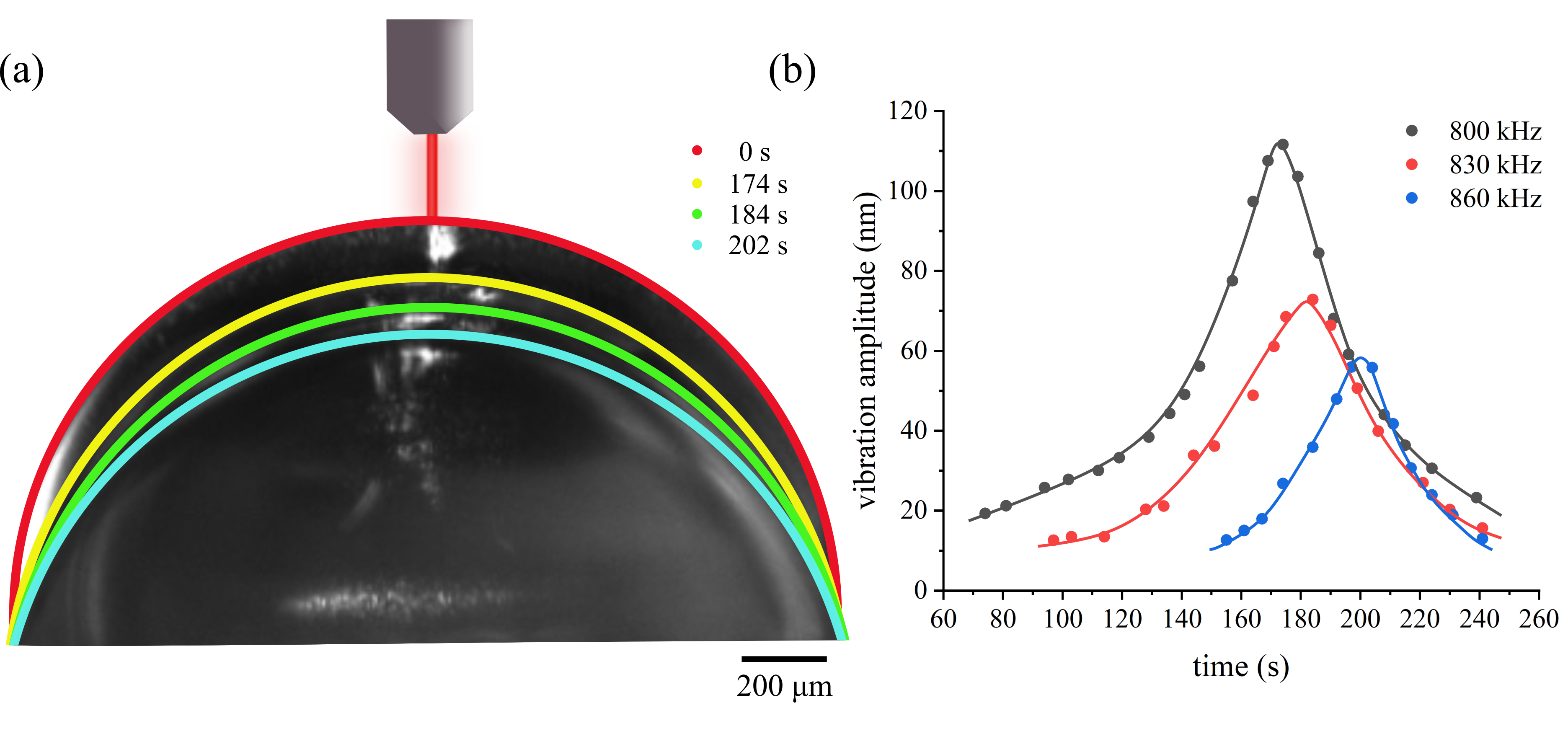}
	\caption{\label{fig:LDV amplitude} Measurements of the droplet surface vibrations by Laser Doppler Vibrometry (LDV). (a) Experimental profiles of the droplet at $t=$174 s, 184 s and 202 s. (b) Vibration amplitude over time with excitation frequencies of 800 kHz, 830 kHz and 860 kHz. }
\end{figure*}

\begin{figure*}
	\includegraphics{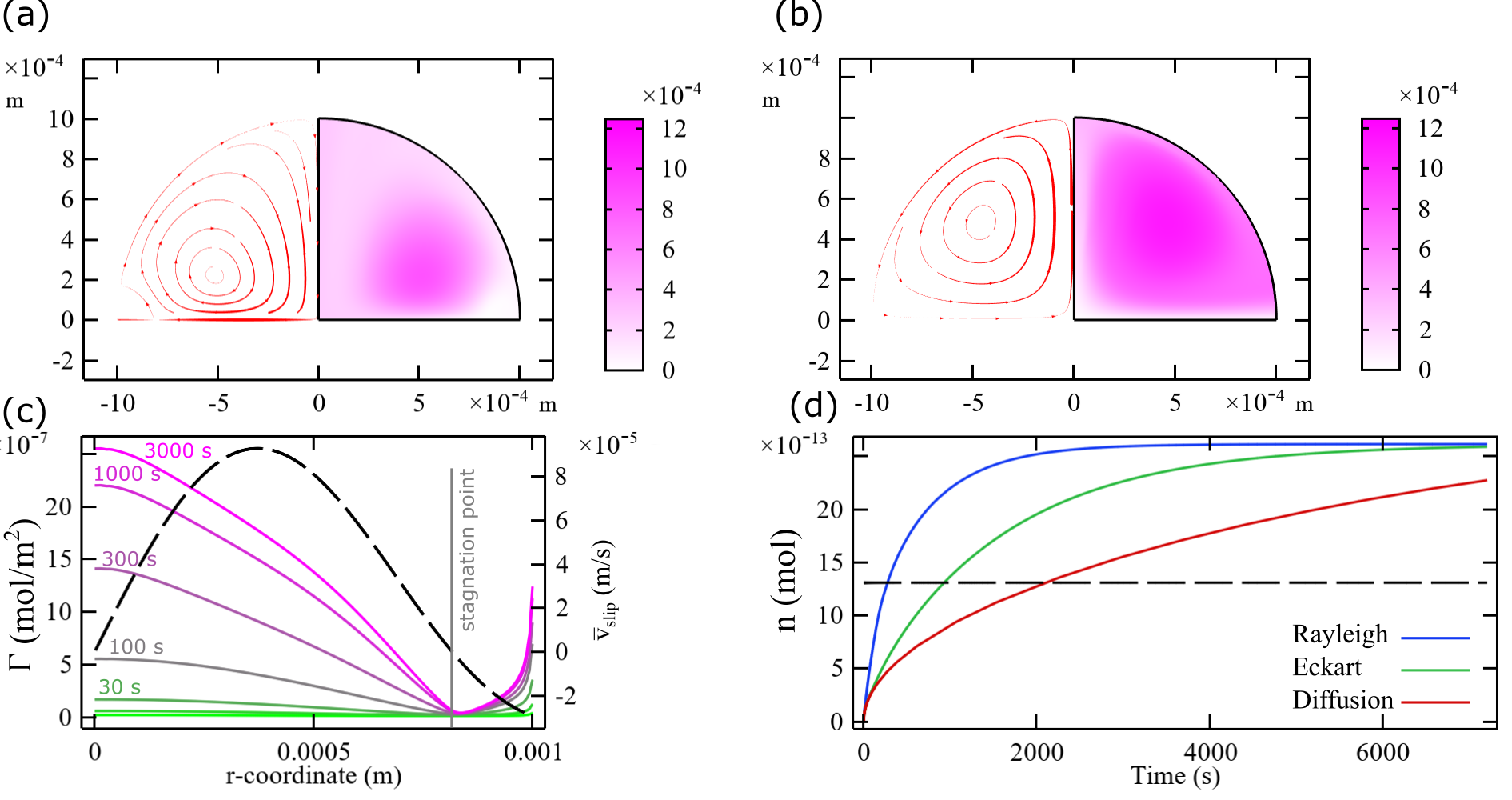}
	\caption{\label{fig:BAW and SAW} Numerical simulation of acoustic mixing in droplets. (a,b) Hydrodynamic flow streamlines (left) and analyte concentration profile (right) at $t=10$ min in Rayleigh (a) and Eckart (b) mixing. The thickness of the streamlines is proportional to the flow velocity. (c) Radial distribution of adsorbed analyte ($\Gamma$) on the biosensor surface at different time points. To locate stagnation points, the slip velocity (\eq{eq: Nyborg}) is given by the right axis. (d) Total amount of adsorbed analyte ($n$) on the sensor surface over time for various mass transfer methods. The dashed line indicates half the amount of analyte initially available.}
\end{figure*}

\begin{figure}
	\includegraphics{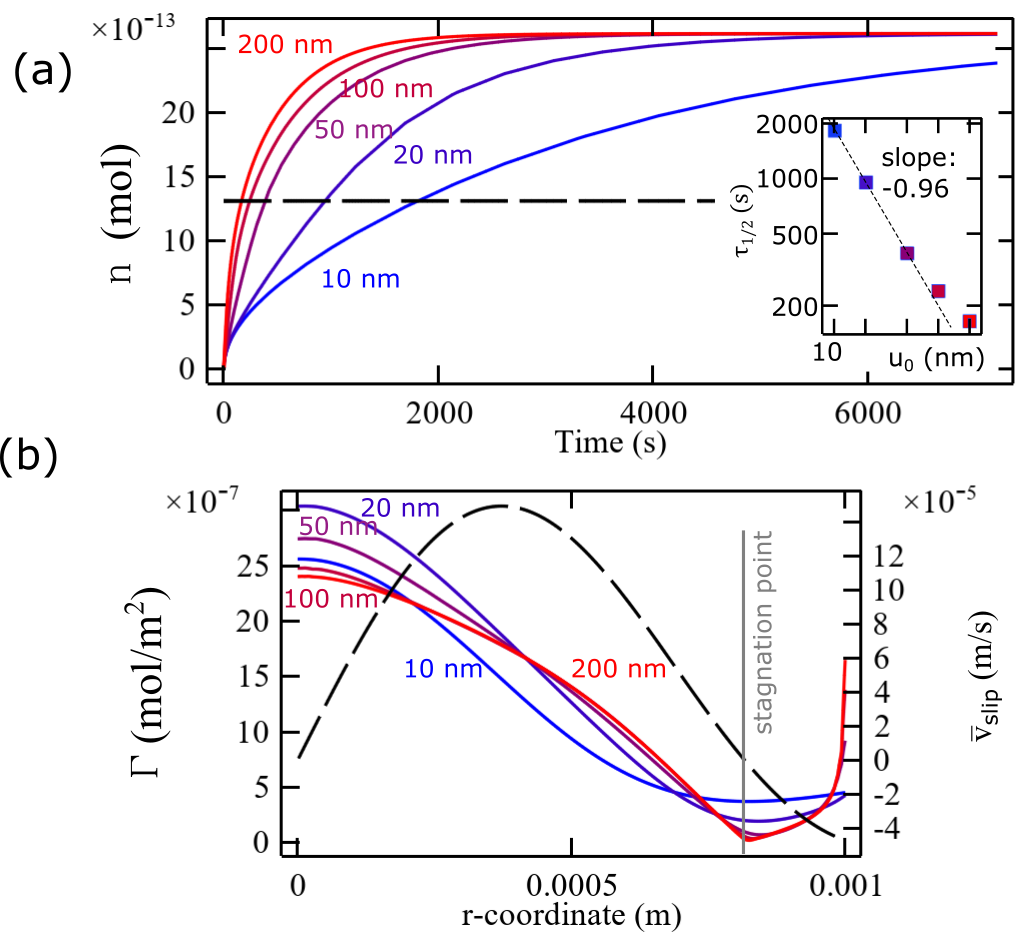}
	\caption{\label{fig:amplitude_sim} (a) Total amount of adsorbed analyte ($n$) on the sensor surface over time for various droplet surface vibration amplitude ($u_0$). The dashed line indicates half the amount of analyte initially available. Inset: scaling of the adsorption characteristic time $\tau_{1/2}$ depending on the droplet vibration amplitude. (b) Effect of the droplet surface vibration amplitude ($u_0$) on the radial distribution of adsorbed analyte ($\Gamma$) on the biosensor surface after 2 h. To locate stagnation points, the slip velocity (\eq{eq: Nyborg}) is given by the right axis.}
\end{figure}

\subsection{Rayleigh acoustic streaming in droplets}
We first optimize acoustic excitation to maximize the acoustic streaming velocity. For water droplets on glass, the velocity of the acoustic streaming is proportional to the ultrasonic energy density in the liquid. In high-frequency Eckart streaming experiments, the acoustic energy is stored in whispering gallery modes that drive a fast acoustic streaming \cite{riaud2017influence}. Provided that the acoustic wavelength is much shorter than the droplet size, these modes are insensitive to the ultrasonic frequency and the exact geometry of the droplet. However, at low frequencies characteristic of Rayleigh streaming, the acoustic wavelength becomes comparable to the droplet and energy can only be stored in discrete frequency modes. Among those modes, \eq{eq: Nyborg} shows that only those having a velocity $\tilde{V}_r$ tangential to the solid walls can drive a strong Rayleigh streaming. In the appendix, we show that the fundamental spherical breathing mode can efficiently excite Rayleigh streaming. At acoustic resonance, the acoustic field intensity peaks, which leads to the strongest acoustic streaming. However, due to evaporation, the droplet volume and contact angle change over time, and so does the resonance frequency. The experimental droplet profiles at three different stages of evaporation are shown as dotted lines in \reffig{fig:LDV amplitude} (a), in which the red profile is the initial state of the droplet. \reffig{fig:LDV amplitude} (b) shows the experimental vibration amplitude of the droplet apex over time for a given excitation frequency. All the curves feature a bell-curve that indicates the onset of acoustic resonance. For instance, a droplet excited at 800 kHz becomes resonant after 174 s of evaporation, whereas a droplet excited at 860 kHz resonates after 202 s. Conversely, it is conceivable that a droplet could be maintained resonant at all times by dynamically adjusting the frequency. 

In order to visualize the acoustic streaming, the three-dimensional trajectory of 2 $\mu$m fluorescent particles dispersed in the droplet is reconstructed by GDPT and the computed velocity field is shown in \reffig{fig:streaming GDPT} (b). The poloidal flow compares qualitatively well to our numerical simulation. The experimental average velocity is 10 $\mu$m/s, which is of similar magnitude to the 16.0 $\mu$m/s obtained from simulations of a droplet vibration amplitude of 78 nm based on vibrometer data (see SI). Discrepancies are most likely due to the experimental excitation failing to hit exactly the resonance frequency. 

\subsection{Acceleration of mass transfer by Rayleigh acoustic streaming}

Having validated the numerical method, we use our model to compare the mass transfer performance of pure diffusion, Eckart streaming and Rayleigh streaming. For the sake of comparison, the average velocity of Eckart streaming is set to the same value as Rayleigh streaming (16.0 $\mu$m/s). The resulting velocity fields and concentration profiles after 5 min are shown in figure \reffig{fig:BAW and SAW}(a,b) (the concentration in the pure diffusion case is shown in SI). Concentration distribution in droplets in each case after 5 min indicates that the analyte is mainly depleted near the solid in the case of Eckart streaming and diffusion, whereas Rayleigh streaming has a much better uniformity. Simulations also reveal that the spatial distribution of adsorbed analyte (adsorption profile) evolves over time \reffig{fig:BAW and SAW}(c). The adsorption is mainly concentrated at the stagnation point at the center of the droplet, where streaming intensity is minimum. This is similar to the case of Rayleigh-streaming-enhanced heat transfer between two plates held at constant temperature where heat transfer primarily occurs at the acoustic velocity nodes \cite{vainshtein1995acoustic}. However, the adsorption minimum is located at another stagnation point. We believe that there is almost no analyte adsorbed there because the inward poloidal flow ensures that all analyte is adsorbed before reaching this second stagnation point. In order to compare the mixing performance of Eckart acoustic streaming, Rayleigh acoustic streaming and pure diffusion \reffig{fig:BAW and SAW}(d), we define $\tau_{1/2}$ as the time it takes to transport half the analyte to the solid surface. Diffusion takes $\tau_{1/2}=2108$ s (35 min) , SAW induced Eckart streaming shortens this process to $\tau_{1/2}=929$ s ($\approx15$ min) and Rayleigh streaming takes $\tau_{1/2}=275$ s ($\approx5$ min). Therefore, Rayleigh streaming is 3 times faster than Eckart mixing. 

Finally, we use the simulations to evaluate the effect of the excitation amplitude on $\tau_{1/2}$  \reffig{fig:amplitude_sim}(a). For small oscillation amplitude ($\leq$ 50 nm), we obtain a linear relationship, similar to Vainshtein \etal \cite{vainshtein1995acoustic}. At high amplitude, the adsorption kinetic differs from the high Peclet asymptotic regime of Vainshtein \etal \cite{vainshtein1995acoustic}. In their work, adsorption is mainly concentrated at the hydrodynamic nodes, whereas in the case of the droplet the streaming stagnation points are more ambiguous and can either be the location of maxima or minima of adsorption.

\begin{figure}
	\includegraphics{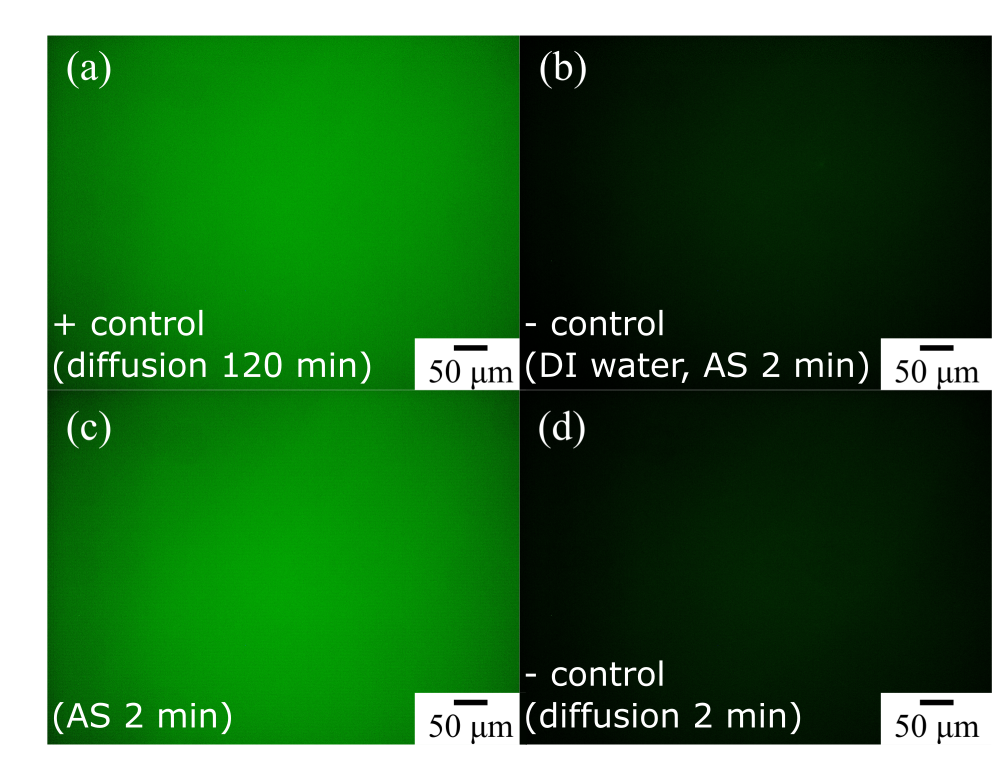}
	\caption{\label{fig:fluorescence compare} Fluorescence microscopy images after incubation in different conditions (a) 0.2 mg/mL FITC $^{*}$ Goat Anti-Mouse lgG (H+L) after two hours of diffusion at 37 $^{\circ}$C, (b) deionized (DI) water after 2 min of Rayleigh acoustic streaming (AS), (c) 0.2 mg/mL FITC $^{*}$ Goat Anti-Mouse lgG (H+L) after 2 min of Rayleigh acoustic streaming (d) 0.2 mg/mL FITC $^{*}$ Goat Anti-Mouse lgG (H+L) after 5 min of diffusion.}
\end{figure}

\begin{figure}
	\includegraphics{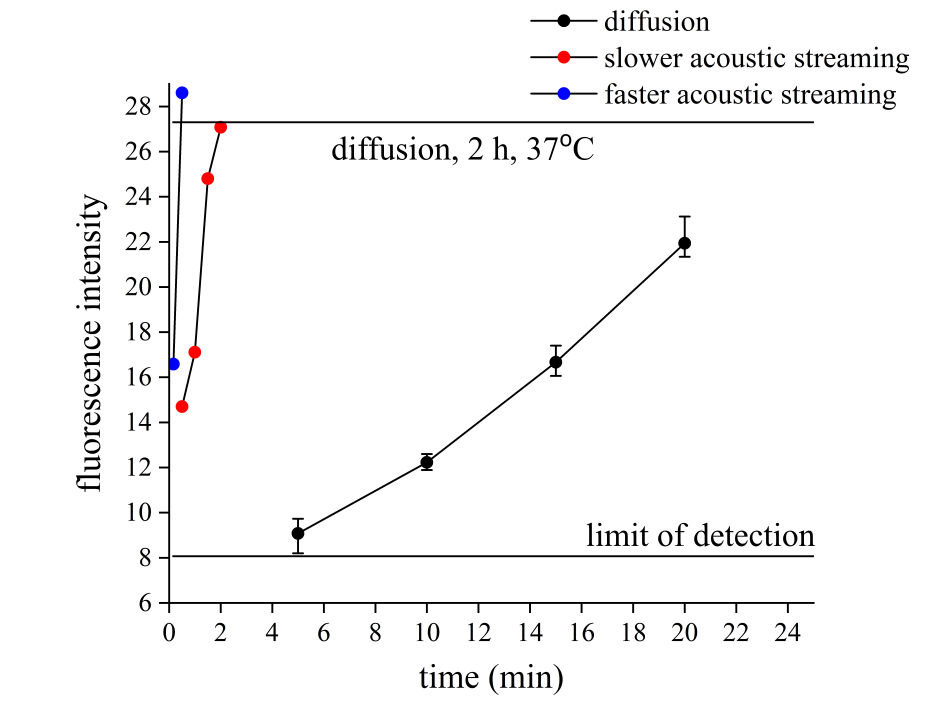}
	\caption{\label{fig:fluorescence intensity} Comparison of adsorption kinetics of 0.2 mg/mL FITC $^{*}$ Goat Anti-Mouse lgG (H+L) with and without acoustic mixing. The limit of detection is obtained from DI water.}
\end{figure}

Next, we evaluate experimentally the acceleration of protein mass transfer to the surface. A fluorescent antibody (FITC $^{*}$ Goat Anti-Mouse lgG (H+L)) able to bind to silanized glass is used to assess mass transfer enhancement due to acoustic streaming. The intensity of fluorescence after different conditions of incubation is shown in Fig.~\reffig{fig:fluorescence compare}. A positive and negative control samples are prepared to bracket the pixel intensity value between the noise floor and the maximum expected intensity. The positive control (\reffig{fig:fluorescence compare} (a)) is the fluorescence obtained after immersing the slide in the antibody solution for 2 h at 37 $^{\circ}$C. The negative control (\reffig{fig:fluorescence compare} (b)) is the fluorescence image of a pure deionized (DI) water sample after 2 min of acoustic streaming. Finally, the fluorescence level after 2 min of antibody adsorption with and without streaming are shown in \reffig{fig:fluorescence compare} (c,d), respectively. The droplet without acoustic mixing looks similar to the negative control (DI water), whereas the droplet that was acoustically mixed looks like the positive control, indicating a saturated adsorption.

The variation of fluorescence intensity over time is shown in \reffig{fig:fluorescence intensity} for pure diffusion and acoustic streaming cases. In the case of pure diffusion, the fluorescence intensity increases steadily, but after 20 minutes the droplet is completely evaporated and yet the fluorescence intensity remains weaker than the positive control (2 h, 37 $^{\circ}$C). In contrast, acoustically mixed droplets reach the saturation value within 30 s (fastest sample) to 2 min (slowest sample), the dispersion being to due the sensitive interplay between droplet volume and resonance frequency (\reffig{fig:LDV amplitude}). 

\subsection{Acceleration of SARS-CoV-2 detection with acoustic streaming}

\begin{figure}
	\includegraphics{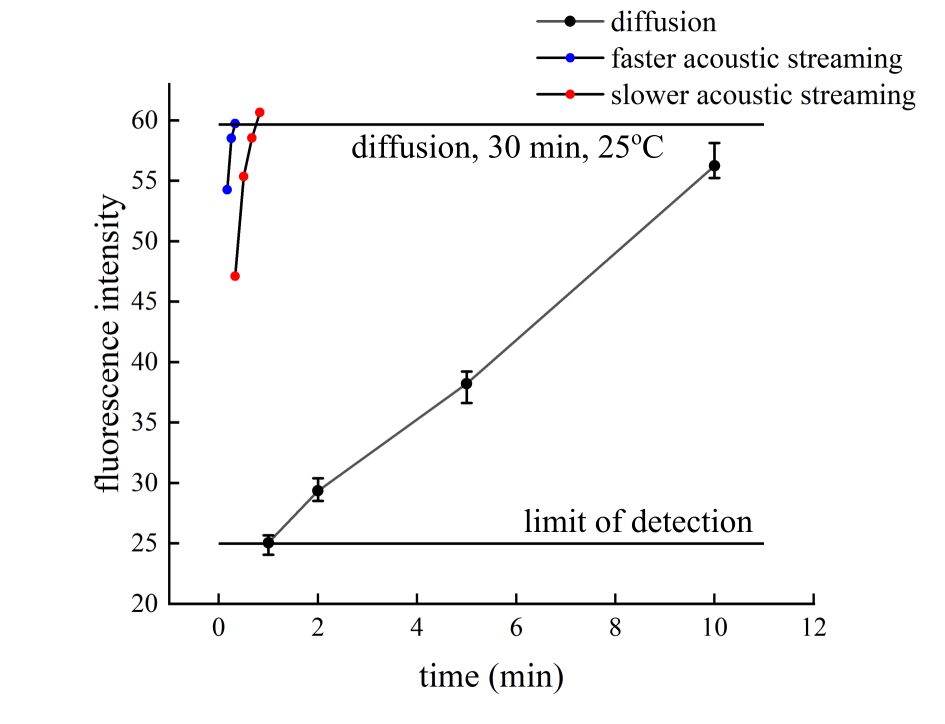}
	\caption{\label{fig:SARS fluorescence intensity} Acceleration of an immunofluorescent SARS-CoV-2 (2019-nCoV) Spike Neutralizing Antibody detection assay. The incubation time for each antibody is kept the same, and is indicated on the horizontal axis. Therefore,  a 10 min value on the axis indicates a total detection duration of 20 min. Limit of detection is obtained from the inflection point of the titer curve (SI).}
\end{figure}

Droplet-based assays are known to reduce reagent use. For surface bioassays, the important metric is the surface concentration of each species, and therefore the total amount of reagent needed is proportional to the surface area of the droplet bottom or, in microplates, of the well bottom. The droplet used in this paper are 2 mm diameter, which is twice smaller than the well of a 96-well plate (5 mm) recommended for such assays. While antigen and secondary antibody concentration were not optimized, we note that 1.5 ng of SARS-CoV-2 Spike Neutralizing Antibody are required for this assay whereas 96-well plate protocols typically recommend 30 to 60 ng per well \cite{dutscher}.

We demonstrate the acceleration of a bioassay to measure the level of SARS-CoV-2 specific antibody (SARS-CoV-2 Spike Neutralizing Antibody) in a liquid sample (here 0.5\% Tween Phosphate Buffer Saline (PBST)). The SARS-CoV-2 specific antibody (primary antibody) specifically binds to an antigen on the surface (2019-nCoV Spike(S) Protein (His-Avi)). This primary antibody is then detected by a fluorescent secondary antibody (FITC $^{*}$ Goat Anti-Mouse lgG (H+L))). Therefore, this experiment involves two incubation steps. For the sake of simplicity, both steps have the same duration.

The  fluorescence levels depending on the incubation method (diffusion or acoustic mixing) are shown in \reffig{fig:SARS fluorescence intensity}. Accordingly, 20 s of acoustic mixing yields fluorescence levels comparable to 10 min of diffusion. This shows that Rayleigh streaming is a highly efficient method to accelerate surface bioassays. It is also instructive to compare Rayleigh streaming results to previous immunoassays using Eckart streaming \cite{bourquin2011integrated,liu2018integrating} that needed 5 to 10 min of mixing (or ``sample concentration'') to complete. Even though these assays did not use the exact same experimental conditions (see Table \ref{tab: speedup}), the speed up offered by Rayleigh streaming is approximately 3 times larger than Eckart mixing, both in simulations and experiments.

\begin{table*}
	\caption{Comparison of adsorption speed up between Eckart and Rayleigh mixing. \label{tab: speedup}}
	\begin{ruledtabular}
		\begin{tabular}{lccccc}
		Method & Droplet volume ($\mu$L) & fluorescent species & acoustic duration & speedup & Ref.\\
		Eckart & ``microliter'' & 2 $\mu$m beads & 5 min 40 s & NA & \cite{bourquin2011integrated}\\
		Eckart & 20 & fluorescent carcinoembryonic antibody & 10 min & 12 & \cite{liu2018integrating}\\
		Rayleigh & 2 & SARS-CoV-2 Spike Neutralizing Antibody & 20 s & 30 & This work (Exp)\\
		Eckart & 2 & NA & 15 min & 2.3 & This work (Sim)\\
		Rayleigh & 2 & NA & 5 min & 7.6 & This work (Sim)\\
		\end{tabular}
	\end{ruledtabular}
\end{table*}

\section{Conclusion}
Accelerating fluid mixing in droplet is a key problem to improve biosensors speed without compromising on reagent consumption. In this paper, we propose to use Rayleigh acoustic streaming to bypass the hydrodynamic boundary layer that has long limited SAW-based droplet mixing. Our simulations and experiments unveil the three-dimensional hydrodynamic flow in the droplet. The simulations also revel the adsorption profile of the analyte in the droplet and the scaling of the adoption with the acoustic excitation amplitude. Experimentally, we demonstrate that Rayleigh streaming can speed up SARS-CoV-2 Spike Neutralizing Antibody detection by up to 30 times, and reduce detection time below 20 s. The application of acoustic streaming in immunoassays provides the possibility for rapid detection of emergent pathogens.

\begin{acknowledgments}
	This work was supported by the National Natural Science Foundation of China with Grant Nos. 12004078 and 61874033; State Key Lab of ASIC and System, Fudan University Nos. 2021MS001, 2021MS002, and 2020KF006; Science and Technology Commission of Shanghai Municipality No. 22QA1400900 and No. 22WZ2502200. This project was supported by the Ministry of Science and Technology of China (Grant No. 2021YFC0863400, 2022YFE0114700), G4 funding from Institut Pasteur, Fondation Merieux and Chinese Academy of Sciences to G.W., and the International Affairs Department of the Institut Pasteur of Paris.
\end{acknowledgments}

\appendix
\section{Acoustic field in the droplet}
The acoustic field in the droplet must satisfy the d'Alembert equation $\Delta \tilde{p}-\frac{1}{{c_0}^2}\partial^2_{tt}\tilde{p}=0$. A well-known solution basis are the spherical harmonics $\tilde{p}_\ell^{m}=p_0 j_\ell(k \varrho)P_\ell^m[\cos(\vartheta)] \cos(\ell\varphi)\cos(\omega t)$, with $k=\omega/c_0$ the wavenumber and $P_\ell^m$ the associated Legendre polynomial of degree $\ell$ and order $m$ ($|m|\leq \ell$). These two indices are determined by the boundary conditions of the acoustic problem. 

In first approximation, the air-liquid interface can be considered infinitely soft $\tilde{p}=0$, and the liquid-solid interface as infinitely hard $\tilde{v}=0$, where the acoustic vibration velocity field $\tilde{v}$ is given by the linearized Euler equation $\rho_0\partial_t\tilde{\mathbf{v}} = -\nabla \tilde{p}$. Therefore, writing $R$ the droplet radius, the acoustic field must satisfy $\left.j_\ell(k \varrho)\right|_{\varrho=R}=0$. This yields a discrete set of allowed values for $k$, the largest one being for $\ell=0$: $kR=\pi$. Setting $\ell=0$ immediately sets $m=0$, and therefore $\tilde{p} = p_0 j_0(k\varrho)$. It is straightforward to show that $\left.\partial_z \tilde{p}\right|_{z=0} = \left.-\frac{1}{\varrho}\partial_\vartheta \tilde{p}\right|_{\vartheta=\pi/2} = 0$. Therefore $\tilde{p} = p_0 j_0(k\varrho)$ is solution of the acoustic problem.

The value of $p_0$ is obtained from the LDV measurements. At the top of the droplet, $\tilde{v}_\text{LDV} = \left.\tilde{v}_z\right|_{z=R} = \left.\tilde{v}_\varrho\right|_{\vartheta=0} = -\frac{1}{\rho_0}\left. \partial_{\varrho}\int_t\tilde{p}dt\right|_{\vartheta=0,\varrho=R}$, this yields $\tilde{v}_\text{LDV} = \frac{k}{\rho_0} p_0 j_1(kR)\sin(\omega t)$. The droplet vibration amplitude at a power of 4.63 W is measured to be 120 nm. The streaming experiments are carried out at 2 W. Since the droplet and substrate are essentially linear (acoustic streaming is a very small perturbation), the displacement scales quadratically with the electrical power, therefore the displacement at 2 W is taken as 78 nm.

\section{Generalization of Nyborg formula for axisymmetric acoustic streaming}
Nyborg calculation is valid for standing waves at solid-fluid interfaces and yields an expression for an effective slip velocity near walls exposed to acoustic waves at a resolution between the viscoacoustic boundary layer thickness and the acoustic wavelength \cite{nyborg1958acoustic}. For standing waves, the pressure may be written $\tilde{p} = \tilde{P}(\mathbf{\varrho})\cos(\omega t)$, and consequently the velocity reads $\tilde{\mathbf{v}} = \tilde{\mathbf{V}}(\mathbf{\varrho})\sin(\omega t)$. In the following, the linearized conservation of mass and the linearized conservation of momentum will be useful:
\begin{eqnarray}
	\partial_t\tilde{p}+{c_0}^2\rho_0\nabla\cdot\tilde{\mathbf{v}} & \text{ reads } & \rho_0{c_0}^2\nabla\cdot\tilde{\mathbf{V}} = \omega\tilde{P} \label{eq: mass cons}\\
	\rho_0\partial_t\tilde{\mathbf{v}} = -\nabla \tilde{p} & \text{ reads } & \rho_0\omega\tilde{\mathbf{V}} = -\nabla \tilde{P} \label{eq: moment cons}
\end{eqnarray}

For a flat hard surface (purely tangential acoustic velocity field), Nyborg slip reads:
\begin{equation}
	\bar{\mathbf{v}}_{slip} = -\left(\frac{1}{4\omega}\right) \left[\mathbf{Q} + 2 \tilde{\mathbf{V}}(\nabla\cdot\tilde{\mathbf{V}}-\partial_z \tilde{V}_z)\right],
	\label{eq: Nyborg flat}
\end{equation}
with $\mathbf{Q} = \tilde{\mathbf{V}}\cdot\nabla\tilde{\mathbf{V}}$. Near the surface, the velocity field is purely radial, which yields $Q_r = \frac{1}{2}\partial_r\tilde{V}_r$.

\eq{eq: mass cons} yields $\nabla\cdot\tilde{\mathbf{V}} = \frac{\omega}{\rho_0{c_0}^2}\tilde{P}$. In \cite{nyborg1958acoustic}, this term was eliminated by the incompressible assumption. Using \eq{eq: moment cons}, the vertical component is recast as $-\partial_z \tilde{V}_z = \frac{1}{\rho_0\omega}\partial^2_{zz} \tilde{P} = \frac{1}{\rho_0\omega}\left[\Delta\tilde{P} - \frac{1}{r}\partial_r r \partial_r \tilde{P} \right] = \frac{1}{\rho_0\omega}\left[-k^2\tilde{P} - \frac{1}{r}\partial_r r \partial_r \tilde{P} \right]$. The $k^2\tilde{P}$ was eliminated by the incompressible assumption in \cite{nyborg1958acoustic}.

Combining all the terms in \eq{eq: Nyborg flat}, we find that $\frac{k^2\tilde{P}}{\rho_0\omega}$ and $\nabla\cdot\tilde{\mathbf{V}}$ that were omitted in the original calculation \cite{nyborg1958acoustic} cancel each-other. All the other terms combined yield \eq{eq: Nyborg}, in accordance with \cite{nyborg1958acoustic}.


\begin{thebibliography}{36}%
	\makeatletter
	\providecommand \@ifxundefined [1]{%
		\@ifx{#1\undefined}
	}%
	\providecommand \@ifnum [1]{%
		\ifnum #1\expandafter \@firstoftwo
		\else \expandafter \@secondoftwo
		\fi
	}%
	\providecommand \@ifx [1]{%
		\ifx #1\expandafter \@firstoftwo
		\else \expandafter \@secondoftwo
		\fi
	}%
	\providecommand \natexlab [1]{#1}%
	\providecommand \enquote  [1]{``#1''}%
	\providecommand \bibnamefont  [1]{#1}%
	\providecommand \bibfnamefont [1]{#1}%
	\providecommand \citenamefont [1]{#1}%
	\providecommand \href@noop [0]{\@secondoftwo}%
	\providecommand \href [0]{\begingroup \@sanitize@url \@href}%
	\providecommand \@href[1]{\@@startlink{#1}\@@href}%
	\providecommand \@@href[1]{\endgroup#1\@@endlink}%
	\providecommand \@sanitize@url [0]{\catcode `\\12\catcode `\$12\catcode
		`\&12\catcode `\#12\catcode `\^12\catcode `\_12\catcode `\%12\relax}%
	\providecommand \@@startlink[1]{}%
	\providecommand \@@endlink[0]{}%
	\providecommand \url  [0]{\begingroup\@sanitize@url \@url }%
	\providecommand \@url [1]{\endgroup\@href {#1}{\urlprefix }}%
	\providecommand \urlprefix  [0]{URL }%
	\providecommand \Eprint [0]{\href }%
	\providecommand \doibase [0]{https://doi.org/}%
	\providecommand \selectlanguage [0]{\@gobble}%
	\providecommand \bibinfo  [0]{\@secondoftwo}%
	\providecommand \bibfield  [0]{\@secondoftwo}%
	\providecommand \translation [1]{[#1]}%
	\providecommand \BibitemOpen [0]{}%
	\providecommand \bibitemStop [0]{}%
	\providecommand \bibitemNoStop [0]{.\EOS\space}%
	\providecommand \EOS [0]{\spacefactor3000\relax}%
	\providecommand \BibitemShut  [1]{\csname bibitem#1\endcsname}%
	\let\auto@bib@innerbib\@empty
	\bibitem [{\citenamefont {Gervais}\ and\ \citenamefont
		{Jensen}(2006)}]{gervais2006mass}%
	\BibitemOpen
	\bibfield  {author} {\bibinfo {author} {\bibfnamefont {T.}~\bibnamefont
			{Gervais}}\ and\ \bibinfo {author} {\bibfnamefont {K.~F.}\ \bibnamefont
			{Jensen}},\ }\bibfield  {title} {\bibinfo {title} {Mass transport and surface
			reactions in microfluidic systems},\ }\href@noop {} {\bibfield  {journal}
		{\bibinfo  {journal} {Chemical engineering science}\ }\textbf {\bibinfo
			{volume} {61}},\ \bibinfo {pages} {1102} (\bibinfo {year}
		{2006})}\BibitemShut {NoStop}%
	\bibitem [{\citenamefont {Hansen}\ \emph {et~al.}(2012)\citenamefont {Hansen},
		\citenamefont {Bruus}, \citenamefont {Callisen},\ and\ \citenamefont
		{Hassager}}]{hansen2012transient}%
	\BibitemOpen
	\bibfield  {author} {\bibinfo {author} {\bibfnamefont {R.}~\bibnamefont
			{Hansen}}, \bibinfo {author} {\bibfnamefont {H.}~\bibnamefont {Bruus}},
		\bibinfo {author} {\bibfnamefont {T.~H.}\ \bibnamefont {Callisen}},\ and\
		\bibinfo {author} {\bibfnamefont {O.}~\bibnamefont {Hassager}},\ }\bibfield
	{title} {\bibinfo {title} {Transient convection, diffusion, and adsorption in
			surface-based biosensors},\ }\href@noop {} {\bibfield  {journal} {\bibinfo
			{journal} {Langmuir}\ }\textbf {\bibinfo {volume} {28}},\ \bibinfo {pages}
		{7557} (\bibinfo {year} {2012})}\BibitemShut {NoStop}%
	\bibitem [{\citenamefont {Pereiro}\ \emph {et~al.}(2020)\citenamefont
		{Pereiro}, \citenamefont {Fomitcheva-Khartchenko},\ and\ \citenamefont
		{Kaigala}}]{pereiro2020shake}%
	\BibitemOpen
	\bibfield  {author} {\bibinfo {author} {\bibfnamefont {I.}~\bibnamefont
			{Pereiro}}, \bibinfo {author} {\bibfnamefont {A.}~\bibnamefont
			{Fomitcheva-Khartchenko}},\ and\ \bibinfo {author} {\bibfnamefont {G.~V.}\
			\bibnamefont {Kaigala}},\ }\bibfield  {title} {\bibinfo {title} {Shake it or
			shrink it: Mass transport and kinetics in surface bioassays using agitation
			and microfluidics},\ }\href@noop {} {\bibfield  {journal} {\bibinfo
			{journal} {Analytical Chemistry}\ }\textbf {\bibinfo {volume} {92}},\
		\bibinfo {pages} {10187} (\bibinfo {year} {2020})}\BibitemShut {NoStop}%
	\bibitem [{\citenamefont {Vanneste}\ and\ \citenamefont
		{B{\"u}hler}(2011)}]{vanneste2011streaming}%
	\BibitemOpen
	\bibfield  {author} {\bibinfo {author} {\bibfnamefont {J.}~\bibnamefont
			{Vanneste}}\ and\ \bibinfo {author} {\bibfnamefont {O.}~\bibnamefont
			{B{\"u}hler}},\ }\bibfield  {title} {\bibinfo {title} {Streaming by leaky
			surface acoustic waves},\ }\href@noop {} {\bibfield  {journal} {\bibinfo
			{journal} {Proceedings of the Royal Society A: Mathematical, Physical and
				Engineering Sciences}\ }\textbf {\bibinfo {volume} {467}},\ \bibinfo {pages}
		{1779} (\bibinfo {year} {2011})}\BibitemShut {NoStop}%
	\bibitem [{\citenamefont {Sadhal}(2012)}]{sadhal2012acoustofluidics}%
	\BibitemOpen
	\bibfield  {author} {\bibinfo {author} {\bibfnamefont {S.}~\bibnamefont
			{Sadhal}},\ }\bibfield  {title} {\bibinfo {title} {Acoustofluidics 15:
			Streaming with sound waves interacting with solid particles},\ }\href@noop {}
	{\bibfield  {journal} {\bibinfo  {journal} {Lab on a Chip}\ }\textbf
		{\bibinfo {volume} {12}},\ \bibinfo {pages} {2600} (\bibinfo {year}
		{2012})}\BibitemShut {NoStop}%
	\bibitem [{\citenamefont {Eckart}(1948)}]{eckart1948vortices}%
	\BibitemOpen
	\bibfield  {author} {\bibinfo {author} {\bibfnamefont {C.}~\bibnamefont
			{Eckart}},\ }\bibfield  {title} {\bibinfo {title} {Vortices and streams
			caused by sound waves},\ }\href@noop {} {\bibfield  {journal} {\bibinfo
			{journal} {Physical review}\ }\textbf {\bibinfo {volume} {73}},\ \bibinfo
		{pages} {68} (\bibinfo {year} {1948})}\BibitemShut {NoStop}%
	\bibitem [{\citenamefont {Rayleigh}(1884)}]{rayleigh1884circulation}%
	\BibitemOpen
	\bibfield  {author} {\bibinfo {author} {\bibfnamefont {L.}~\bibnamefont
			{Rayleigh}},\ }\bibfield  {title} {\bibinfo {title} {On the circulation of
			air observed in kundt's tubes, and on some allied acoustical problems},\
	}\href@noop {} {\bibfield  {journal} {\bibinfo  {journal} {Philosophical
				Transactions of the Royal Society of London}\ }\textbf {\bibinfo {volume}
			{175}},\ \bibinfo {pages} {1} (\bibinfo {year} {1884})}\BibitemShut {NoStop}%
	\bibitem [{\citenamefont {Nyborg}(1958)}]{nyborg1958acoustic}%
	\BibitemOpen
	\bibfield  {author} {\bibinfo {author} {\bibfnamefont {W.~L.}\ \bibnamefont
			{Nyborg}},\ }\bibfield  {title} {\bibinfo {title} {Acoustic streaming near a
			boundary},\ }\href@noop {} {\bibfield  {journal} {\bibinfo  {journal} {The
				Journal of the Acoustical Society of America}\ }\textbf {\bibinfo {volume}
			{30}},\ \bibinfo {pages} {329} (\bibinfo {year} {1958})}\BibitemShut
	{NoStop}%
	\bibitem [{\citenamefont {Bach}\ and\ \citenamefont
		{Bruus}(2018)}]{bach2018theory}%
	\BibitemOpen
	\bibfield  {author} {\bibinfo {author} {\bibfnamefont {J.~S.}\ \bibnamefont
			{Bach}}\ and\ \bibinfo {author} {\bibfnamefont {H.}~\bibnamefont {Bruus}},\
	}\bibfield  {title} {\bibinfo {title} {Theory of pressure acoustics with
			viscous boundary layers and streaming in curved elastic cavities},\
	}\href@noop {} {\bibfield  {journal} {\bibinfo  {journal} {The Journal of the
				Acoustical Society of America}\ }\textbf {\bibinfo {volume} {144}},\ \bibinfo
		{pages} {766} (\bibinfo {year} {2018})}\BibitemShut {NoStop}%
	\bibitem [{\citenamefont {Shilton}\ \emph {et~al.}(2008)\citenamefont
		{Shilton}, \citenamefont {Tan}, \citenamefont {Yeo},\ and\ \citenamefont
		{Friend}}]{shilton2008particle}%
	\BibitemOpen
	\bibfield  {author} {\bibinfo {author} {\bibfnamefont {R.}~\bibnamefont
			{Shilton}}, \bibinfo {author} {\bibfnamefont {M.~K.}\ \bibnamefont {Tan}},
		\bibinfo {author} {\bibfnamefont {L.~Y.}\ \bibnamefont {Yeo}},\ and\ \bibinfo
		{author} {\bibfnamefont {J.~R.}\ \bibnamefont {Friend}},\ }\bibfield  {title}
	{\bibinfo {title} {Particle concentration and mixing in microdrops driven by
			focused surface acoustic waves},\ }\href@noop {} {\bibfield  {journal}
		{\bibinfo  {journal} {Journal of Applied Physics}\ }\textbf {\bibinfo
			{volume} {104}},\ \bibinfo {pages} {014910} (\bibinfo {year}
		{2008})}\BibitemShut {NoStop}%
	\bibitem [{\citenamefont {Wixforth}(2003)}]{wixforth2003acoustically}%
	\BibitemOpen
	\bibfield  {author} {\bibinfo {author} {\bibfnamefont {A.}~\bibnamefont
			{Wixforth}},\ }\bibfield  {title} {\bibinfo {title} {Acoustically driven
			planar microfluidics},\ }\href@noop {} {\bibfield  {journal} {\bibinfo
			{journal} {Superlattices and Microstructures}\ }\textbf {\bibinfo {volume}
			{33}},\ \bibinfo {pages} {389} (\bibinfo {year} {2003})}\BibitemShut
	{NoStop}%
	\bibitem [{\citenamefont {Bourquin}\ \emph {et~al.}(2011)\citenamefont
		{Bourquin}, \citenamefont {Reboud}, \citenamefont {Wilson}, \citenamefont
		{Zhang},\ and\ \citenamefont {Cooper}}]{bourquin2011integrated}%
	\BibitemOpen
	\bibfield  {author} {\bibinfo {author} {\bibfnamefont {Y.}~\bibnamefont
			{Bourquin}}, \bibinfo {author} {\bibfnamefont {J.}~\bibnamefont {Reboud}},
		\bibinfo {author} {\bibfnamefont {R.}~\bibnamefont {Wilson}}, \bibinfo
		{author} {\bibfnamefont {Y.}~\bibnamefont {Zhang}},\ and\ \bibinfo {author}
		{\bibfnamefont {J.~M.}\ \bibnamefont {Cooper}},\ }\bibfield  {title}
	{\bibinfo {title} {Integrated immunoassay using tuneable surface acoustic
			waves and lensfree detection},\ }\href@noop {} {\bibfield  {journal}
		{\bibinfo  {journal} {Lab on a Chip}\ }\textbf {\bibinfo {volume} {11}},\
		\bibinfo {pages} {2725} (\bibinfo {year} {2011})}\BibitemShut {NoStop}%
	\bibitem [{\citenamefont {Renaudin}\ \emph {et~al.}(2010)\citenamefont
		{Renaudin}, \citenamefont {Chabot}, \citenamefont {Grondin}, \citenamefont
		{Aimez},\ and\ \citenamefont {Charette}}]{renaudin2010integrated}%
	\BibitemOpen
	\bibfield  {author} {\bibinfo {author} {\bibfnamefont {A.}~\bibnamefont
			{Renaudin}}, \bibinfo {author} {\bibfnamefont {V.}~\bibnamefont {Chabot}},
		\bibinfo {author} {\bibfnamefont {E.}~\bibnamefont {Grondin}}, \bibinfo
		{author} {\bibfnamefont {V.}~\bibnamefont {Aimez}},\ and\ \bibinfo {author}
		{\bibfnamefont {P.~G.}\ \bibnamefont {Charette}},\ }\bibfield  {title}
	{\bibinfo {title} {Integrated active mixing and biosensing using surface
			acoustic waves (saw) and surface plasmon resonance (spr) on a common
			substrate},\ }\href@noop {} {\bibfield  {journal} {\bibinfo  {journal} {Lab
				on a Chip}\ }\textbf {\bibinfo {volume} {10}},\ \bibinfo {pages} {111}
		(\bibinfo {year} {2010})}\BibitemShut {NoStop}%
	\bibitem [{\citenamefont {Salman}\ and\ \citenamefont
		{Haynes}(2007)}]{salman2007numerical}%
	\BibitemOpen
	\bibfield  {author} {\bibinfo {author} {\bibfnamefont {H.}~\bibnamefont
			{Salman}}\ and\ \bibinfo {author} {\bibfnamefont {P.}~\bibnamefont
			{Haynes}},\ }\bibfield  {title} {\bibinfo {title} {A numerical study of
			passive scalar evolution in peripheral regions},\ }\href@noop {} {\bibfield
		{journal} {\bibinfo  {journal} {Physics of Fluids}\ }\textbf {\bibinfo
			{volume} {19}},\ \bibinfo {pages} {067101} (\bibinfo {year}
		{2007})}\BibitemShut {NoStop}%
	\bibitem [{\citenamefont {Lebedev}\ and\ \citenamefont
		{Turitsyn}(2004)}]{lebedev2004passive}%
	\BibitemOpen
	\bibfield  {author} {\bibinfo {author} {\bibfnamefont {V.}~\bibnamefont
			{Lebedev}}\ and\ \bibinfo {author} {\bibfnamefont {K.}~\bibnamefont
			{Turitsyn}},\ }\bibfield  {title} {\bibinfo {title} {Passive scalar evolution
			in peripheral regions},\ }\href@noop {} {\bibfield  {journal} {\bibinfo
			{journal} {Physical Review E}\ }\textbf {\bibinfo {volume} {69}},\ \bibinfo
		{pages} {036301} (\bibinfo {year} {2004})}\BibitemShut {NoStop}%
	\bibitem [{\citenamefont {Li}\ \emph {et~al.}(2019)\citenamefont {Li},
		\citenamefont {Huffman}, \citenamefont {Ranganathan}, \citenamefont {He},\
		and\ \citenamefont {Li}}]{li2019acoustofluidic}%
	\BibitemOpen
	\bibfield  {author} {\bibinfo {author} {\bibfnamefont {X.}~\bibnamefont
			{Li}}, \bibinfo {author} {\bibfnamefont {J.}~\bibnamefont {Huffman}},
		\bibinfo {author} {\bibfnamefont {N.}~\bibnamefont {Ranganathan}}, \bibinfo
		{author} {\bibfnamefont {Z.}~\bibnamefont {He}},\ and\ \bibinfo {author}
		{\bibfnamefont {P.}~\bibnamefont {Li}},\ }\bibfield  {title} {\bibinfo
		{title} {Acoustofluidic enzyme-linked immunosorbent assay (elisa) platform
			enabled by coupled acoustic streaming},\ }\href@noop {} {\bibfield  {journal}
		{\bibinfo  {journal} {Analytica Chimica Acta}\ }\textbf {\bibinfo {volume}
			{1079}},\ \bibinfo {pages} {129} (\bibinfo {year} {2019})}\BibitemShut
	{NoStop}%
	\bibitem [{\citenamefont {Zhang}\ \emph {et~al.}(2021)\citenamefont {Zhang},
		\citenamefont {Brunet}, \citenamefont {Royon},\ and\ \citenamefont
		{Guo}}]{zhang2021mixing}%
	\BibitemOpen
	\bibfield  {author} {\bibinfo {author} {\bibfnamefont {C.}~\bibnamefont
			{Zhang}}, \bibinfo {author} {\bibfnamefont {P.}~\bibnamefont {Brunet}},
		\bibinfo {author} {\bibfnamefont {L.}~\bibnamefont {Royon}},\ and\ \bibinfo
		{author} {\bibfnamefont {X.}~\bibnamefont {Guo}},\ }\bibfield  {title}
	{\bibinfo {title} {Mixing intensification using sound-driven micromixer with
			sharp edges},\ }\href@noop {} {\bibfield  {journal} {\bibinfo  {journal}
			{Chemical Engineering Journal}\ }\textbf {\bibinfo {volume} {410}},\ \bibinfo
		{pages} {128252} (\bibinfo {year} {2021})}\BibitemShut {NoStop}%
	\bibitem [{\citenamefont {Rasouli}\ and\ \citenamefont
		{Tabrizian}(2019)}]{rasouli2019ultra}%
	\BibitemOpen
	\bibfield  {author} {\bibinfo {author} {\bibfnamefont {M.~R.}\ \bibnamefont
			{Rasouli}}\ and\ \bibinfo {author} {\bibfnamefont {M.}~\bibnamefont
			{Tabrizian}},\ }\bibfield  {title} {\bibinfo {title} {An ultra-rapid acoustic
			micromixer for synthesis of organic nanoparticles},\ }\href@noop {}
	{\bibfield  {journal} {\bibinfo  {journal} {Lab on a Chip}\ }\textbf
		{\bibinfo {volume} {19}},\ \bibinfo {pages} {3316} (\bibinfo {year}
		{2019})}\BibitemShut {NoStop}%
	\bibitem [{\citenamefont {Chen}\ \emph {et~al.}(2018)\citenamefont {Chen},
		\citenamefont {Chen}, \citenamefont {Bai}, \citenamefont {Gao}, \citenamefont
		{Metcalfe}, \citenamefont {Cheng},\ and\ \citenamefont
		{Zhu}}]{chen2018multiplexed}%
	\BibitemOpen
	\bibfield  {author} {\bibinfo {author} {\bibfnamefont {H.}~\bibnamefont
			{Chen}}, \bibinfo {author} {\bibfnamefont {C.}~\bibnamefont {Chen}}, \bibinfo
		{author} {\bibfnamefont {S.}~\bibnamefont {Bai}}, \bibinfo {author}
		{\bibfnamefont {Y.}~\bibnamefont {Gao}}, \bibinfo {author} {\bibfnamefont
			{G.}~\bibnamefont {Metcalfe}}, \bibinfo {author} {\bibfnamefont
			{W.}~\bibnamefont {Cheng}},\ and\ \bibinfo {author} {\bibfnamefont
			{Y.}~\bibnamefont {Zhu}},\ }\bibfield  {title} {\bibinfo {title} {Multiplexed
			detection of cancer biomarkers using a microfluidic platform integrating
			single bead trapping and acoustic mixing techniques},\ }\href@noop {}
	{\bibfield  {journal} {\bibinfo  {journal} {Nanoscale}\ }\textbf {\bibinfo
			{volume} {10}},\ \bibinfo {pages} {20196} (\bibinfo {year}
		{2018})}\BibitemShut {NoStop}%
	\bibitem [{\citenamefont {Meng}\ \emph {et~al.}(2019)\citenamefont {Meng},
		\citenamefont {Liu}, \citenamefont {Wang}, \citenamefont {Zhang},
		\citenamefont {Zhou}, \citenamefont {Cai}, \citenamefont {Li}, \citenamefont
		{Wu}, \citenamefont {Xu}, \citenamefont {Niu} \emph
		{et~al.}}]{meng2019sonoporation}%
	\BibitemOpen
	\bibfield  {author} {\bibinfo {author} {\bibfnamefont {L.}~\bibnamefont
			{Meng}}, \bibinfo {author} {\bibfnamefont {X.}~\bibnamefont {Liu}}, \bibinfo
		{author} {\bibfnamefont {Y.}~\bibnamefont {Wang}}, \bibinfo {author}
		{\bibfnamefont {W.}~\bibnamefont {Zhang}}, \bibinfo {author} {\bibfnamefont
			{W.}~\bibnamefont {Zhou}}, \bibinfo {author} {\bibfnamefont {F.}~\bibnamefont
			{Cai}}, \bibinfo {author} {\bibfnamefont {F.}~\bibnamefont {Li}}, \bibinfo
		{author} {\bibfnamefont {J.}~\bibnamefont {Wu}}, \bibinfo {author}
		{\bibfnamefont {L.}~\bibnamefont {Xu}}, \bibinfo {author} {\bibfnamefont
			{L.}~\bibnamefont {Niu}}, \emph {et~al.},\ }\bibfield  {title} {\bibinfo
		{title} {Sonoporation of cells by a parallel stable cavitation microbubble
			array},\ }\href@noop {} {\bibfield  {journal} {\bibinfo  {journal} {Advanced
				Science}\ }\textbf {\bibinfo {volume} {6}},\ \bibinfo {pages} {1900557}
		(\bibinfo {year} {2019})}\BibitemShut {NoStop}%
	\bibitem [{\citenamefont {Marin}\ \emph {et~al.}(2015)\citenamefont {Marin},
		\citenamefont {Rossi}, \citenamefont {Rallabandi}, \citenamefont {Wang},
		\citenamefont {Hilgenfeldt},\ and\ \citenamefont
		{K{\"a}hler}}]{marin2015three}%
	\BibitemOpen
	\bibfield  {author} {\bibinfo {author} {\bibfnamefont {A.}~\bibnamefont
			{Marin}}, \bibinfo {author} {\bibfnamefont {M.}~\bibnamefont {Rossi}},
		\bibinfo {author} {\bibfnamefont {B.}~\bibnamefont {Rallabandi}}, \bibinfo
		{author} {\bibfnamefont {C.}~\bibnamefont {Wang}}, \bibinfo {author}
		{\bibfnamefont {S.}~\bibnamefont {Hilgenfeldt}},\ and\ \bibinfo {author}
		{\bibfnamefont {C.~J.}\ \bibnamefont {K{\"a}hler}},\ }\bibfield  {title}
	{\bibinfo {title} {Three-dimensional phenomena in microbubble acoustic
			streaming},\ }\href@noop {} {\bibfield  {journal} {\bibinfo  {journal}
			{Physical Review Applied}\ }\textbf {\bibinfo {volume} {3}},\ \bibinfo
		{pages} {041001} (\bibinfo {year} {2015})}\BibitemShut {NoStop}%
	\bibitem [{\citenamefont {Bengtsson}\ and\ \citenamefont
		{Laurell}(2004)}]{bengtsson2004ultrasonic}%
	\BibitemOpen
	\bibfield  {author} {\bibinfo {author} {\bibfnamefont {M.}~\bibnamefont
			{Bengtsson}}\ and\ \bibinfo {author} {\bibfnamefont {T.}~\bibnamefont
			{Laurell}},\ }\bibfield  {title} {\bibinfo {title} {Ultrasonic agitation in
			microchannels},\ }\href@noop {} {\bibfield  {journal} {\bibinfo  {journal}
			{Analytical and bioanalytical chemistry}\ }\textbf {\bibinfo {volume}
			{378}},\ \bibinfo {pages} {1716} (\bibinfo {year} {2004})}\BibitemShut
	{NoStop}%
	\bibitem [{\citenamefont {Royer}\ and\ \citenamefont
		{Dieulesaint}(1986)}]{royer1986optical}%
	\BibitemOpen
	\bibfield  {author} {\bibinfo {author} {\bibfnamefont {D.}~\bibnamefont
			{Royer}}\ and\ \bibinfo {author} {\bibfnamefont {E.}~\bibnamefont
			{Dieulesaint}},\ }\bibfield  {title} {\bibinfo {title} {Optical probing of
			the mechanical impulse response of a transducer},\ }\href@noop {} {\bibfield
		{journal} {\bibinfo  {journal} {Applied physics letters}\ }\textbf {\bibinfo
			{volume} {49}},\ \bibinfo {pages} {1056} (\bibinfo {year}
		{1986})}\BibitemShut {NoStop}%
	\bibitem [{\citenamefont {Muller}\ \emph {et~al.}(2013)\citenamefont {Muller},
		\citenamefont {Rossi}, \citenamefont {Marin}, \citenamefont {Barnkob},
		\citenamefont {Augustsson}, \citenamefont {Laurell}, \citenamefont
		{Kaehler},\ and\ \citenamefont {Bruus}}]{muller2013ultrasound}%
	\BibitemOpen
	\bibfield  {author} {\bibinfo {author} {\bibfnamefont {P.~B.}\ \bibnamefont
			{Muller}}, \bibinfo {author} {\bibfnamefont {M.}~\bibnamefont {Rossi}},
		\bibinfo {author} {\bibfnamefont {A.}~\bibnamefont {Marin}}, \bibinfo
		{author} {\bibfnamefont {R.}~\bibnamefont {Barnkob}}, \bibinfo {author}
		{\bibfnamefont {P.}~\bibnamefont {Augustsson}}, \bibinfo {author}
		{\bibfnamefont {T.}~\bibnamefont {Laurell}}, \bibinfo {author} {\bibfnamefont
			{C.~J.}\ \bibnamefont {Kaehler}},\ and\ \bibinfo {author} {\bibfnamefont
			{H.}~\bibnamefont {Bruus}},\ }\bibfield  {title} {\bibinfo {title}
		{Ultrasound-induced acoustophoretic motion of microparticles in three
			dimensions},\ }\href@noop {} {\bibfield  {journal} {\bibinfo  {journal}
			{Physical Review E}\ }\textbf {\bibinfo {volume} {88}},\ \bibinfo {pages}
		{023006} (\bibinfo {year} {2013})}\BibitemShut {NoStop}%
	\bibitem [{\citenamefont {Barnkob}\ \emph {et~al.}(2015)\citenamefont
		{Barnkob}, \citenamefont {K{\"a}hler},\ and\ \citenamefont
		{Rossi}}]{barnkob2015general}%
	\BibitemOpen
	\bibfield  {author} {\bibinfo {author} {\bibfnamefont {R.}~\bibnamefont
			{Barnkob}}, \bibinfo {author} {\bibfnamefont {C.~J.}\ \bibnamefont
			{K{\"a}hler}},\ and\ \bibinfo {author} {\bibfnamefont {M.}~\bibnamefont
			{Rossi}},\ }\bibfield  {title} {\bibinfo {title} {General defocusing particle
			tracking},\ }\href@noop {} {\bibfield  {journal} {\bibinfo  {journal} {Lab on
				a Chip}\ }\textbf {\bibinfo {volume} {15}},\ \bibinfo {pages} {3556}
		(\bibinfo {year} {2015})}\BibitemShut {NoStop}%
	\bibitem [{\citenamefont {Barnkob}\ and\ \citenamefont
		{Rossi}(2021)}]{barnkob2021defocustracker}%
	\BibitemOpen
	\bibfield  {author} {\bibinfo {author} {\bibfnamefont {R.}~\bibnamefont
			{Barnkob}}\ and\ \bibinfo {author} {\bibfnamefont {M.}~\bibnamefont
			{Rossi}},\ }\bibfield  {title} {\bibinfo {title} {Defocustracker: A modular
			toolbox for defocusing-based, single-camera, 3d particle tracking},\
	}\href@noop {} {\bibfield  {journal} {\bibinfo  {journal} {arXiv preprint
				arXiv:2102.03056}\ } (\bibinfo {year} {2021})}\BibitemShut {NoStop}%
	\bibitem [{\citenamefont {Rossi}\ \emph {et~al.}(2019)\citenamefont {Rossi},
		\citenamefont {Marin},\ and\ \citenamefont
		{K{\"a}hler}}]{rossi2019interfacial}%
	\BibitemOpen
	\bibfield  {author} {\bibinfo {author} {\bibfnamefont {M.}~\bibnamefont
			{Rossi}}, \bibinfo {author} {\bibfnamefont {A.}~\bibnamefont {Marin}},\ and\
		\bibinfo {author} {\bibfnamefont {C.~J.}\ \bibnamefont {K{\"a}hler}},\
	}\bibfield  {title} {\bibinfo {title} {Interfacial flows in sessile
			evaporating droplets of mineral water},\ }\href@noop {} {\bibfield  {journal}
		{\bibinfo  {journal} {Physical Review E}\ }\textbf {\bibinfo {volume}
			{100}},\ \bibinfo {pages} {033103} (\bibinfo {year} {2019})}\BibitemShut
	{NoStop}%
	\bibitem [{\citenamefont {Shiokawa}\ \emph {et~al.}(1990)\citenamefont
		{Shiokawa}, \citenamefont {Matsui},\ and\ \citenamefont
		{Ueda}}]{shiokawa1990study}%
	\BibitemOpen
	\bibfield  {author} {\bibinfo {author} {\bibfnamefont {S.}~\bibnamefont
			{Shiokawa}}, \bibinfo {author} {\bibfnamefont {Y.}~\bibnamefont {Matsui}},\
		and\ \bibinfo {author} {\bibfnamefont {T.}~\bibnamefont {Ueda}},\ }\bibfield
	{title} {\bibinfo {title} {Study on saw streaming and its application to
			fluid devices},\ }\href@noop {} {\bibfield  {journal} {\bibinfo  {journal}
			{Japanese journal of applied physics}\ }\textbf {\bibinfo {volume} {29}},\
		\bibinfo {pages} {137} (\bibinfo {year} {1990})}\BibitemShut {NoStop}%
	\bibitem [{\citenamefont {Raghavan}\ \emph {et~al.}(2010)\citenamefont
		{Raghavan}, \citenamefont {Friend},\ and\ \citenamefont
		{Yeo}}]{raghavan2010particle}%
	\BibitemOpen
	\bibfield  {author} {\bibinfo {author} {\bibfnamefont {R.~V.}\ \bibnamefont
			{Raghavan}}, \bibinfo {author} {\bibfnamefont {J.~R.}\ \bibnamefont
			{Friend}},\ and\ \bibinfo {author} {\bibfnamefont {L.~Y.}\ \bibnamefont
			{Yeo}},\ }\bibfield  {title} {\bibinfo {title} {Particle concentration via
			acoustically driven microcentrifugation: micropiv flow visualization and
			numerical modelling studies},\ }\href@noop {} {\bibfield  {journal} {\bibinfo
			{journal} {Microfluidics and Nanofluidics}\ }\textbf {\bibinfo {volume}
			{8}},\ \bibinfo {pages} {73} (\bibinfo {year} {2010})}\BibitemShut {NoStop}%
	\bibitem [{\citenamefont {Riaud}\ \emph {et~al.}(2017)\citenamefont {Riaud},
		\citenamefont {Baudoin}, \citenamefont {Bou~Matar}, \citenamefont {Thomas},\
		and\ \citenamefont {Brunet}}]{riaud2017influence}%
	\BibitemOpen
	\bibfield  {author} {\bibinfo {author} {\bibfnamefont {A.}~\bibnamefont
			{Riaud}}, \bibinfo {author} {\bibfnamefont {M.}~\bibnamefont {Baudoin}},
		\bibinfo {author} {\bibfnamefont {O.}~\bibnamefont {Bou~Matar}}, \bibinfo
		{author} {\bibfnamefont {J.-L.}\ \bibnamefont {Thomas}},\ and\ \bibinfo
		{author} {\bibfnamefont {P.}~\bibnamefont {Brunet}},\ }\bibfield  {title}
	{\bibinfo {title} {On the influence of viscosity and caustics on acoustic
			streaming in sessile droplets: an experimental and a numerical study with a
			cost-effective method},\ }\href@noop {} {\bibfield  {journal} {\bibinfo
			{journal} {Journal of Fluid Mechanics}\ }\textbf {\bibinfo {volume} {821}},\
		\bibinfo {pages} {384} (\bibinfo {year} {2017})}\BibitemShut {NoStop}%
	\bibitem [{\citenamefont {Peng}\ \emph {et~al.}(2022)\citenamefont {Peng},
		\citenamefont {Li}, \citenamefont {Zhou},\ and\ \citenamefont
		{Jiang}}]{peng2022concentration}%
	\BibitemOpen
	\bibfield  {author} {\bibinfo {author} {\bibfnamefont {T.}~\bibnamefont
			{Peng}}, \bibinfo {author} {\bibfnamefont {L.}~\bibnamefont {Li}}, \bibinfo
		{author} {\bibfnamefont {M.}~\bibnamefont {Zhou}},\ and\ \bibinfo {author}
		{\bibfnamefont {F.}~\bibnamefont {Jiang}},\ }\bibfield  {title} {\bibinfo
		{title} {Concentration of microparticles using flexural acoustic wave in
			sessile droplets},\ }\href@noop {} {\bibfield  {journal} {\bibinfo  {journal}
			{Sensors}\ }\textbf {\bibinfo {volume} {22}},\ \bibinfo {pages} {1269}
		(\bibinfo {year} {2022})}\BibitemShut {NoStop}%
	\bibitem [{\citenamefont {Muller}\ \emph {et~al.}(2012)\citenamefont {Muller},
		\citenamefont {Barnkob}, \citenamefont {Jensen},\ and\ \citenamefont
		{Bruus}}]{muller2012numerical}%
	\BibitemOpen
	\bibfield  {author} {\bibinfo {author} {\bibfnamefont {P.~B.}\ \bibnamefont
			{Muller}}, \bibinfo {author} {\bibfnamefont {R.}~\bibnamefont {Barnkob}},
		\bibinfo {author} {\bibfnamefont {M.~J.~H.}\ \bibnamefont {Jensen}},\ and\
		\bibinfo {author} {\bibfnamefont {H.}~\bibnamefont {Bruus}},\ }\bibfield
	{title} {\bibinfo {title} {A numerical study of microparticle acoustophoresis
			driven by acoustic radiation forces and streaming-induced drag forces},\
	}\href@noop {} {\bibfield  {journal} {\bibinfo  {journal} {Lab on a Chip}\
		}\textbf {\bibinfo {volume} {12}},\ \bibinfo {pages} {4617} (\bibinfo {year}
		{2012})}\BibitemShut {NoStop}%
	\bibitem [{\citenamefont {Pokri{\'c}}\ and\ \citenamefont
		{Pu{\v{c}}ar}(1979)}]{pokric1979two}%
	\BibitemOpen
	\bibfield  {author} {\bibinfo {author} {\bibfnamefont {B.}~\bibnamefont
			{Pokri{\'c}}}\ and\ \bibinfo {author} {\bibfnamefont {Z.}~\bibnamefont
			{Pu{\v{c}}ar}},\ }\bibfield  {title} {\bibinfo {title} {The two-cross
			immunodiffusion technique: diffusion coefficients and precipitating titers of
			igg in human serum and rabbit serum antibodies},\ }\href@noop {} {\bibfield
		{journal} {\bibinfo  {journal} {Analytical biochemistry}\ }\textbf {\bibinfo
			{volume} {93}},\ \bibinfo {pages} {103} (\bibinfo {year} {1979})}\BibitemShut
	{NoStop}%
	\bibitem [{\citenamefont {Vainshtein}\ \emph {et~al.}(1995)\citenamefont
		{Vainshtein}, \citenamefont {Fichman},\ and\ \citenamefont
		{Gutfinger}}]{vainshtein1995acoustic}%
	\BibitemOpen
	\bibfield  {author} {\bibinfo {author} {\bibfnamefont {P.}~\bibnamefont
			{Vainshtein}}, \bibinfo {author} {\bibfnamefont {M.}~\bibnamefont
			{Fichman}},\ and\ \bibinfo {author} {\bibfnamefont {C.}~\bibnamefont
			{Gutfinger}},\ }\bibfield  {title} {\bibinfo {title} {Acoustic enhancement of
			heat transfer between two parallel plates},\ }\href@noop {} {\bibfield
		{journal} {\bibinfo  {journal} {International Journal of Heat and Mass
				Transfer}\ }\textbf {\bibinfo {volume} {38}},\ \bibinfo {pages} {1893}
		(\bibinfo {year} {1995})}\BibitemShut {NoStop}%
	\bibitem [{dut()}]{dutscher}%
	\BibitemOpen
	\href@noop {} {\bibinfo {title} {Microplates for enzyme linked immunosorbent
			assays (elisa)}},\ \bibinfo {howpublished}
	{\url{https://www.dutscher.com/data/pdf_guides/en/Guide_de_selection_ELISA_Microlon_Greiner_Bio-One.pdf}},\
	\bibinfo {note} {accessed: 2022-08-31}\BibitemShut {NoStop}%
	\bibitem [{\citenamefont {Liu}\ \emph {et~al.}(2018)\citenamefont {Liu},
		\citenamefont {Li},\ and\ \citenamefont
		{Bhethanabotla}}]{liu2018integrating}%
	\BibitemOpen
	\bibfield  {author} {\bibinfo {author} {\bibfnamefont {J.}~\bibnamefont
			{Liu}}, \bibinfo {author} {\bibfnamefont {S.}~\bibnamefont {Li}},\ and\
		\bibinfo {author} {\bibfnamefont {V.~R.}\ \bibnamefont {Bhethanabotla}},\
	}\bibfield  {title} {\bibinfo {title} {Integrating metal-enhanced
			fluorescence and surface acoustic waves for sensitive and rapid
			quantification of cancer biomarkers from real matrices},\ }\href@noop {}
	{\bibfield  {journal} {\bibinfo  {journal} {ACS sensors}\ }\textbf {\bibinfo
			{volume} {3}},\ \bibinfo {pages} {222} (\bibinfo {year} {2018})}\BibitemShut
	{NoStop}%
\end{thebibliography}
\providecommand{\noopsort}[1]{}\providecommand{\singleletter}[1]{#1}%

\end{document}